\documentclass{article}

\begin{document}

\title{Spin Glasses: Old and New Complexity
\thanks{This paper is based on a talk given at the Interdisciplinary
Symposium on Complex Systems, Halkidiki, Greece, September 19-25, 2011.}}

\author{D.L.~Stein\thanks{daniel.stein@nyu.edu}\\
Department of Physics and Courant Institute of Mathematical Sciences\\ 
New York University\\
New York, NY 10003 USA\\
\and
C.M.~Newman\thanks{newman@cims.nyu.edu}\\
Courant Institute of Mathematical Sciences, New York University\\
New York, NY 10012 USA
}

\maketitle


\begin{abstract}
Spin glasses are disordered magnetic systems that exhibit a variety of
properties that are characteristic of complex systems.  After a brief
review of basic spin glass concepts, their use in areas such as
computer science, biology, and other fields will be explored.
This use and its underlying basis will be termed old
complexity.  Newer concepts and
ideas flowing from more recent studies of spin glasses will then be discussed, leading
to a proposal for a kind of new complexity.
\end{abstract}

\section{Introduction}
\label{sec:intro}

Spin glasses are disordered magnetic systems in which two nearby localized
magnetic moments have a roughly equal probability of interacting either
through a ferromagnetic interaction (in which energy is lowered by the
moments aligning) or an antiferromagnetic one (energy lowered by
antialigning). This can be achieved in several ways.  The most common is to
substitute, at random locations, a very small concentration (at most a few
percent) of a magnetic element, often iron or manganese, inside a
nonmagnetic metallic host, such as one of the noble metals (copper, silver,
or gold). In these dilute magnetic alloys, spins localized on the impurity
atoms polarize the surrounding conduction electron gas in concentric
spheres (roughly speaking) of alternating spin polarization (this is known
as the RKKY interaction~\cite{AM76}). Depending on the placement of two
nearby magnetic impurities, the conduction-mediated magnetic
interaction can therefore be either ferromagnetic or antiferromagnetic.

But there's more than one way to make a spin glass. For example, europium
strontium sulfide (Eu$_x$Sr$_{1-x}$S, with $x$ typically a few tenths) is
also a spin glass. Here the mechanism is different: nearest-neighbor
interactions are ferromagnetic, while next-nearest neighbor are
antiferromagnetic. Because the magnetic impurity europium is substituted
randomly for nonmagnetic strontium, the net effect is again to generate 
competing ferromagnetic and antiferromagnetic interactions. 

The first mechanism leads to a metal, the second an insulator. In addition,
spin glasses can be uniaxial (i.e., spins can point only along one axis) or
isotropic; they can be crystalline or amorphous. Clearly, spin glasses come
in all varieties. What then are the features that they do share?

The most basic is that all possess a disordered ground state configuration:
while a typical low-energy atomic configuration of a glass lacks
long-range translational order, the spin configuration of a spin glass
lacks long-range orientational order; hence the nomenclature.

The signature laboratory features of all spin glasses include~\cite{BY86} a
cusp in the low-field ac~susceptibility at a frequency-dependent
temperature $T_f$; a smoothly rounded maximum in specific heat at a
temperature slightly above $T_f$; localized magnetic moments frozen into
random orientations; and very long relaxational or equilibrational
timescales.  

So why are they interesting?

There are several reasons. From the point of view of physics, spin glasses
and other systems with quenched disorder represent a serious gap in our
understanding of condensed matter.  In more conventional systems,
long-range order and useful symmetries enable us to use many of the
powerful tools of condensed matter physics and statistical mechanics to
establish a conceptual framework to understand their nature and behavior.
Such notions include those of Bloch waves, broken symmetry, order
parameters, Goldstone modes, topological singularities in order-parameter
space, and many others.  There are far fewer tools at our disposal for
understanding systems with quenched disorder, such as glasses or spin
glasses. In this regard, spin glasses are especially interesting because,
unlike ordinary glasses, which must be cooled sufficiently rapidly to avoid
the crystalline phase, the spin glass has no competing ordered phase. So if
a thermodynamic phase transition does exist, then the low temperature phase
would truly be an equilibrium condensed disordered phase -- a new state of
matter.

Note that important qualifier: if a thermodynamic phase transition
exists. It is a remarkable fact that, forty years since spin glasses were
first identified as such~\cite{CMB71}, this most basic of all issues
remains unresolved.  The susceptibility data show a cusp at a temperature
identified as $T_f$, the ``freezing temperature'', indicating a phase
transition; but the specific heat data show only a rounded maximum,
occurring at a temperature typically 10-20\% above $T_f$.  Many experiments
performed since have failed to reconcile these conflicting
results. Numerical simulations have yielded some information, indicating
the existence of a phase transition in three and higher
dimensions~\cite{Ogielski85,OM85,KY96}.  But all we know (from theoretical
work) for certain is that there is no phase transition in one dimension and
that there is one in infinite dimensions.  Everything in between remains
conjectural.

While this is one of the most fundamental questions we can ask about any
condensed matter system, it is only one of many that remain unanswered
after decades of experimental, theoretical, and numerical
investigations. The problem is that spin glasses do not possess the kinds
of symmetries that make accessible the study of crystals, ferromagnets,
superconductors, and other homogeneous systems; the absence of these
symmetries enormously complicates the analysis of spin glass behavior. The
simultaneous presence of both disorder and frustration can lead to new
types of broken symmetries, a breakdown of the thermodynamic limit for
certain quantities, the emergence of new phenomena such as chaotic
temperature dependence, the need for creation of new thermodynamic tools,
and other unanticipated features.  While it may not be necessary to
completely revamp statistical mechanics in order to understand disordered
systems, as has sometimes been suggested, it is at least necessary to
carefully rethink some deeply held assumptions.

Finally, the study of spin glasses has led to a surprising variety of
applications to problems in biology, computer science, economics, and other
areas. We'll briefly mention a few of these applications below, but here
will simply note that the usefulness of spin glass concepts, enabling them
to serve as a bridge to fields outside of physics, is one of the early
reasons why spin glasses came to be regarded as relevant to the study of
complexity.

\section{Formulation of the Problem}
\label{sec:theory}

The modern theory of spin glasses began with the work of Edwards and
Anderson~(EA)~\cite{EA75}, who proposed that the essential physics of spin
glasses lay not in the details of their microscopic interactions but rather
in the competition between quenched ferromagnetic and
antiferromagnetic interactions.  They proposed the Hamiltonian
\begin{equation}
\label{eq:EA}
{\cal H}_{\cal J}=-\sum_{<x,y>} J_{xy} \sigma_x\sigma_y -h\sum_x\sigma_x\ ,
\end{equation}
where $x$ is a site in a $d$-dimensional cubic lattice, $\sigma_x$ is the
spin at site $x$, $h$ is an external magnetic field, and the first sum is
over nearest neighbor pairs of sites only.  The couplings $J_{xy}$ are independent
random variables chosen from a common distribution (such as Gaussian with
mean zero and variance one), and the notation ${\cal J}$ indicates a
particular realization of the couplings, corresponding physically
to a specific spin glass sample.  We hereafter restrict ourselves to Ising
models, where the only allowed spin values are $\sigma_x=\pm 1$.

The disorder is represented by the $J_{xy}$'s and is quenched; once
chosen, they remain fixed for all time, and the spins must adjust as best
they can. Physically, this corresponds to the fact that localized magnetic
moments in laboratory spin glasses (for example, dilute magnetic alloys)
are attached to their host impurity atoms, which do not diffuse on
laboratory -- or indeed, much longer --- timescales.

In reality, no laboratory spin glass has an energy function that looks
like~(\ref{eq:EA}). The great insight behind the EA Hamiltonian is that it
is conjectured to be the simplest Hamiltonian that accurately models real
spin glasses. The essential ingredient is quenched, randomly placed
ferromagnetic and antiferromagnetic couplings between nearby spins.
Given this, one can just as well study the spins on a regular lattice; in a
real host material, the prime effect of the random placing of magnetic
impurities is to generate both ferromagnetic and antiferromagnetic
couplings. The restriction in~(\ref{eq:EA}) to spin-spin interactions
between nearest neighbors only also doesn't occur in real spin glass
materials; but the hope is that this again is more of a detail that does
not alter its applicability. (Since we're very far from solving even the
simple-looking EA Hamiltonian, these assertions have to remain conjectures
for now.)

An immediate, and nontrivial, consequence of the competition between
ferromagnetic and antiferromagnetic interactions in~(\ref{eq:EA}) is the
presence of frustration: no spin configuration can simultaneously
satisfy all couplings. How does one then find the ground state?  Which
couplings should be chosen to be unsatisfied? Or could it be that there are
possibly many ground states --- or at positive temperature,
thermodynamic pure states --- not connected by any simple symmetry
transformation?  This is a very intriguing question, and remains one of the
central unsolved problems in spin glass research.  It is also one of the
prime features of spin glasses that has caught the attention and interest
of complexity researchers.

Before proceeding further, we emphasize that one can have quenched disorder
without frustration (the most well-known example being the Mattis
model~\cite{Mattis76}), and on the flip side, one can have frustration
without disorder (for example, a planar antiferromagnet on a triangular
lattice). Spin glasses are hard to analyze at least partly because of the
joint presence of quenched disorder and frustration --- but this also makes
them useful as a model system with which to examine certain aspects of
complexity, as we'll see.

\section{Mean Field Theory}
\label{sec:mft}

The most studied and best understood spin glass model is the
infinite-ranged version of the EA~Hamiltonian, proposed by Sherrington and
Kirkpatrick~(SK)~\cite{SK75}.  For a system of $N$ Ising~spins in zero
external field, the SK~Hamiltonian is given by

\begin{equation}
\label{eq:SK}
{\cal H}_{{\cal J},N}=-{1\over\sqrt{N}}\sum_{1\le i<j\le N}J_{ij} \sigma_i\sigma_j 
\end{equation}
where the independent, identically distributed couplings $J_{ij}$ are again
chosen from a Gaussian distribution with zero mean and variance one.
Unlike the EA~model, which has $O(N)$ couplings for a system of $N$ spins,
the SK model has $O(N^2)$ couplings; this requires a $1/\sqrt{N}$ rescaling
of the coupling magnitudes to ensure a sensible thermodynamic limit for
free energy per spin and other thermodynamic quantities.

Sherrington and Kirkpatrick showed that their model had an equilibrium
phase transition, but their solution for the low-temperature phase was
unstable~\cite{SK75}.  The correct solution for the low-temperature phase
of the SK~model was found several years later by Parisi~\cite{Parisi79},
who proposed an extraordinary new kind of symmetry breaking, known today as
``replica symmetry breaking'', or RSB.  The essential idea is that the
low-temperature phase consists not of a single spin-reversed pair of
states, but rather of ``infinitely many pure thermodynamic
states''~\cite{Parisi83}, not related by any simple symmetry
transformations.

But the most striking feature of RSB is not the existence of many
non-symmetry-related thermodynamic states --- though this is already very
unusual. What really generated a huge amount of interest and excitement was
the way that the states were organized; in particular, the joint properties
of ``non-self-averaging'' of the distribution of spin overlaps between
thermodynamic states, and an ultrametric distance relation between triples
of pure states~\cite{Parisi83,MPSTV84a,MPSTV84b}.

These require some explanation. We already indicated that there are many
equilibrium spin glass states, and since all are disordered, there are no
easily distinguishable features to categorize them. An alternative is then
to see how similar they are to each other, by introducing a spin overlap function
$q_{\alpha\beta}$ between two pure states $\alpha$ and $\beta$:
\begin{equation}
\label{eq:qab}
q_{\alpha\beta}=\frac{1}{N}\sum_{i=1}^N\langle\sigma_i\rangle_\alpha\langle\sigma_i\rangle_\beta\,
  .
\end{equation}

Although there are many pure thermodynamic states, any spin glass at some
fixed temperature and field can be found in only one, and so we need to
consider the probability $W_\alpha$ that the system will be found in state
$\alpha$ (this of course depends on temperature and field, as well as
${\cal J}$, but we'll suppress these dependences for ease of
notation). $W_\alpha$ is usually called the weight of state
$\alpha$, and of course $\sum_\alpha W_\alpha=1$. Consequently, if you
choose two spin configurations independently from the Gibbs
measure, the probability $P_{\cal J}(q)dq$ that their spin overlap will be
between $q$ and $q+dq$ is given by
\begin{equation}
\label{eq:sod}
P_{\cal J}(q) =
\sum_{\alpha\beta}W_\alpha W_\beta\delta(q-q_{\alpha\beta})\, ,
\end{equation}
where $\delta$ is the Dirac delta-function and $P_{\cal J}(q)$ is called
the spin overlap density.  The subscript ${\cal J}$ indicates that
the procedure is done for fixed coupling realization.

In the SK spin glass, the overlap density has a complicated form. Lengthy
analyses~\cite{MPSTV84a,MPSTV84b,DT85,MPV85,MPV87} have shown that only a
handful of thermodynamic states at low (but nonzero) temperature have
weights that aren't extremely small. One must also take into account the
self-overlap of each pure state with itself; it was shown that all
pure states have the same self-overlap, regardless of coupling realization
(as always, aside from a set of measure zero). The fixed value of this
self-overlap is known as the Edwards-Anderson order parameter $q_{EA}$.
Between $-q_{EA}$ and $+q_{EA}$ one has a complicated structure of overlaps between
different pure states.

But that's not what's surprising. What's surprising is that it was found
that no matter how large $N$ becomes, the positions and weights of all
overlaps strictly between $\pm q_{EA}$ vary with coupling realization.
(The overlaps $\pm q_{EA}$, on the other hand, are the same for all
coupling realizations.) 

This may seem problematic: thermodynamics works because different samples
behave the same way in the large-$N$ limit. Here's a situation where
sample-to-sample fluctuations of an important macroscopic property do 
not diminish as $N\to\infty$. This is what is known as 
non-self-averaging; it represents an important and distinguishing
feature of the SK spin glass. 

Although non-self-averaging is unusual, it can be reconciled with our
understanding of how thermodynamics should behave. The usual, measurable
thermodynamic quantities --- energy, free energy, magnetization (in nonzero
field), and so on --- are in fact self-averaging in the spin glass. The
spin overlap, though important as a theoretical tool for understanding spin
glass order, has no immediate or obvious observable consequences.

What happens if we average the overlap distribution over all
coupling realizations? Analytically we have 
\begin{equation}
\label{eq:Pq}
P(q)=\prod_{1\le i<j\le N}\int dJ_{ij} P(J_{ij}) P_{\cal J}(q)
\end{equation}
also known as the (averaged) Parisi order parameter. Because there are an
uncountable number of coupling realizations, and no special overlap value
(aside from the self-overlap $q_{EA}$ and its negative) that might either
appear in no coupling realization (leading to a gap) or else appear in some
positive-measure set of realizations (leading to a spike), $P(q)$ is smooth
between the two spikes appearing at $\pm q_{EA}$.

This is strikingly different from anything seen in more conventional
homogeneous systems. But there's yet another surprise. Suppose we look at
overlaps from triples of states rather than from a single pair. If we do, we find
that any three states have overlap relations of a very
special kind.  This is usually stated in terms of the ``distance'' between
the states, which is just the overlap subtracted from $q_{EA}$, so the more
dissimilar two states --- i.e., the smaller the overlap --- the larger
their distance in configuration space. (And of course, a state has zero
distance from itself.)

It turns out that the three distances from any three states form the
sides of an equilateral or acute isosceles triangle. A space where this
distance relation always holds among any three points is known as an 
ultrametric space. The canonical example of an ultrametric space is a
nested (or tree-like, or hierarchical) structure.

All this is much different from anything observed in homogeneous systems.
Now, mean-field theory usually provides a reliable description of the
low-temperature properties of finite-dimensional models (and becomes exact above some sufficiently
high but still finite dimension); in particular it is often used to reveal the
nature of the broken symmetry in more difficult to solve finite-dimensional models. 
So it was natural to expect that the RSB
mean-field picture should similarly describe the nature of ordering in the
EA and other short-range spin glass models. This generated a lot of
excitement; in particular, could these properties provide a ``universality
class'' for a wide range of disordered systems -- in particular laboratory
spin glasses and even structural glasses? We'll return to this question
shortly, but turn now to what we might call the ``old complexity'' features
of spin glasses.

\section{``Old'' Complexity}
\label{sec:old}

By the mid-1980's, a number of features of spin glasses brought them to the
attention of scientists interested in problems that had come to be known as
``complex systems''. These spin glass features included their signature
properties of possessing both quenched disorder and frustration; and in
particular, the ease of precisely formulating these properties in the spin
glass context. All complex systems must exhibit these properties in one
form or another; certainly a strict homogeneity or rigid ordering (as in a
crystal) would preclude any chance of evolving or adapting to changing
environments. And without conflicting constraints and requirements, it is
difficult to see how anything approaching complexity --- with its implied
storage and generation of large information content --- could develop.

The spin glass feature that has probably done the most to catch the
attention and interest of complexity scientists, though, is the presence of
many metastable states, that is, states stable to flips of finite numbers
of spins. This is often picturesquely characterized by a ``rugged energy
landscape'' (sometimes also used to denote the presence of many pure or
ground states, known to occur in the SK~model but whose presence remains
controversial in more realistic models such as EA). This feature of many
``near-optimal'' solutions, along with higher-energy (or cost) dynamical
traps, is a feature shared by many problems in complexity.

It is also significant that a hierarchical ordering emerges in the state
space structure of states in the SK model. This type of structure, or
``near-decomposability'', was proposed by Simon~\cite{Simon62,Simon73} as a
universal feature underlying the architecture of complex systems.  The fact
that such a structure emerges naturally from the relatively featureless SK
Hamiltonian is profoundly surprising. Whether this structure arises in more
realistic models, however, is unknown, and its significance for complexity
more generally remains unclear.

All of the features discussed above are static, equilibrium properties; and
moreover are mostly derived from theoretical constructs. But real spin
glasses in the laboratory exhibit nonequilibrium dynamical features that
might be equally, if not more, relevant to other complex systems. That is,
their dynamical behaviors are highly anomalous~\cite{BY86}, including such
characteristics as slow relaxation, irreversibility, memory effects,
hysteresis, and aging. All of these are features shared --- in one form or
another --- by other complex systems.

Many of these features, whether static or dynamical, have lent themselves
to the development of new approaches to studying problems in computer
science, biology, economics, and elsewhere~\cite{MPV87,Stein89b,Stein92};
the construction of new algorithms and computational schema for hard
combinatorial optimization problems; new analytical approaches to finding
bounds on costs of near-optimal solutions in NP-complete problems; models
for protein folding and conformational dynamics, maturation of the immune
response, prebiotic and biological evolution, and neural networks; and new
methods of neural-based computation. The ``bridge''~\cite{Anderson88} that
spin glasses provided to numerous problems outside of physics and
mathematics led to their inclusion in systems of interest to complexity
scientists.

\section{``New'' Complexity?}
\label{sec:new}

Recent work on the structure of short-range spin glasses (which remain
poorly understood) suggests that spin glasses may be ``complex'' in more
subtle ways~\cite{Stein04,NSbook}. The low-temperature thermodynamic
structure of SK spin glasses provides a rich lode of phenomena that has had
a significant impact on complexity studies --- what we referred to here and
in~\cite{NSbook} as ``old'' complexity.  But alternative theoretical
models~\cite{Mac84,FH86,BM87} and mathematical
investigations~\cite{NS96b,NSBerlin,NS98,NS03b} suggest that short-range
spin glasses do not possess mean-field behavior in any finite
dimension (so that the $d\to\infty$ limit is singular) --- though the issue
remains controversial. But if true, then spin glasses would present a
unique statistical mechanical example of such a phenomenon.

Equally importantly, they could help us understand the limits of
applicability of analogies between different types of systems. Complexity
science thrives on analogy, and the resulting transference of concepts and
techniques from one field to another. But true complexity should also bump
up against the limits of this program; at some point every complex system
has to display unique features of some kind, and at a fundamental
level. The concept of universality classes, which has been one of the
central and unifying ideas of statistical mechanics, may be less than well
suited for complex systems. The (possibly) wildly different behaviors
between short-range spin glasses in any dimension and
infinite-range spin glasses (which mimic EA models in infinite dimension)
--- where one would certainly expect similar low-temperature ordering ---
could be a foreshadowing of this sort of {\it sui generis\/} behavior.

As another example, consider the lack of a straightforward thermodynamic
limit for Gibbs states in systems with many competing pure
states~\cite{NS92} (as spin glasses are often claimed to possess).  A core
assumption throughout statistical mechanics is that the thermodynamic limit
reveals the bulk properties of large finite systems. But suppose that the
low-temperature EA spin glass possesses many pure states; in that case the
connection between the thermodynamic limit and the behavior of finite
macroscopic systems becomes far less direct. In fact, the difference
between the thermodynamic behavior of conventional ordered systems and spin
glasses would bear a similar relation to the difference between the
dynamical behavior of a classical system with a single fixed point and one
with a strange attractor.

Spin glasses possess numerous other features that might be relevant in
thinking about other complex systems but that haven't yet been explored in
this context. Such features include the presence of disorder
chaos~\cite{KK07,Chatterjee09}, temperature chaos~\cite{BM87}, and
stochastic stability~\cite{AC98}. Finally, the construction of new
mathematical tools like the metastate~\cite{NSBerlin,NS03b,AW90,NS96c,NS97}
may prove as useful for the study of complex systems as any of the other
new concepts and tools that have arisen from the study of spin glasses.

\section*{Acknowledgments}
This work was supported in part by NSF Grants
DMS-0604869 and DMS-1106316.

\bibliographystyle{ieeetr}
\bibliography{refs}

\end{document}